\documentclass[12pt]{article}
\usepackage{epsf}  
\begin{document}

\newcommand{\sheptitle}
{Numerical consistency check between two approaches to radiative corrections for neutrino masses and mixings}
\newcommand{\shepauthor}
{Mrinal Kumar Das$^{a}$, Mahadev Patgiri$^{b}$  and N.Nimai Singh$^{a,c,}${\footnote{Regular Associate of ICTP; {\it {e-mail address:}} nimai03@yahoo.com}}}

\newcommand{\shepaddress}
{${^a}$Department of Physics, Gauhati University, Guwahati-781014, India \\
${^b}$Department of Physics, Cotton College, Guwahati -781001, India\\
${^c}$ International Centre for Theoretical Physics, Strada Costiera 11,\\
31014 Trieste, Italy 
}
\newcommand{\shepabstract}
{We briefly outline the two popular   approaches on  radiative corrections to neutrino masses and mixing angles, and then 
carry out a detailed  numerical analysis for a consistency check between them in  MSSM. We find that the two approaches are nearly 
consistent with a discrepancy of a factor of $13\%$ in mass eigenvalues at low energy scale but the predictions on mixing angles are almost consistent.
We check the stability of the three types of neutrino models, i.e., 
 hierarchical, inverted hierarchical and degenerate models, under radiative corrections, using 
 both  approaches,  and find consistent conclusions. The neutrino mass models which are found to be stable under radiative corrections in MSSM  
are the normal hierarchical  model and  the   inverted hierarchical model with opposite CP parity. We also carry out numerical analysis on 
some important conjectures 
related to  radiative corrections in MSSM, viz., radiative magnification of solar and atmospheric mixings in case of nearly degenerate model having 
same CP parity (MPR conjecture) 
and radiative generation of solar mass scale in exactly two-fold degenerate model with  opposite CP parity and non-zero $U_{e3}$ (JM conjecture).
 We observe certain exceptions to these conjectures.
 Finally the effect of
 scale-dependent vacuum expectation value  in neutrino mass renormalisation is discussed.}
\begin{titlepage}
\begin{flushright}
hep-ph/0407185
\end{flushright}
\begin{center}
{\large{\bf\sheptitle}}
\bigskip\\
\shepauthor
\\
\mbox{}\\
{\it\shepaddress}\\
\vspace{.5in}
{\bf Abstract}
\bigskip
\end{center}
\setcounter{page}{0}
\shepabstract
\end{titlepage}
\section{Introduction}
Recent developments in  the determination of neutrino masses and mixing angles from various oscillation experiments, 
have strengthened  our knowledge of neutrino physics[1]. In order to have a meaningful comparison of the theoretical
 predictions on neutrino masses and mixing angles  within the framework of GUTs with or without supersymmetry, 
 with the  datas from various neutrino oscillation experiments[2], the effects of radiative corrections are very essential[3].
 Considerable progress has been achieved  at  this front and this can be mainly  classified into two categories -  (i) 
the evolution of the renormalisation group equations (RGEs) from high to low energy scale[4,5,6,7], and 
 (ii) the low energy threshold corrections[8]. In case 
of running the  RGEs in (i), the general  underlying motivations are :
to check the stability of the neutrino mass model under   radiative corrections[9,10];
to generate solar mass scale and also  reactor angle $|U_{e3}|$ from radiative corrections[11];
to generate  correct radiative magnifications of solar and atmospheric mixing angles from CKM-like small values at high scale[12]; and 
to get  suitable deviations  from the bimaximal solar and atmospheric mixings through radiative corrections [13,14], etc.
One basic difference of the first one from the last three points is the very definition of the stability criteria under radiative corrections[15].
Radiative stability generally means that  the effects of radiative corrections  do 
not alter substantially the good predictions on neutrino masses and mixings already acquired through seesaw mechanism  at high  scale $M_R$.  

Within the framework of running the RGEs from high energy scale to low energy scale, we have again two different  approaches so far employed in the literature. 
In the first approach (we call `Method A' for simplicity) the running is carried out through the neutrino mass matrix $m_{LL}$ as a whole, 
and at every energy scale 
one can extract neutrino masses and mixing angles through the diagonalisation of the neutrino mass matrix calculated  at that  particular energy  scale[6,7,10]. 
In the second approach (we call 'Method B') the running of the RGEs is carried out in terms neutrino mass eigenvalues and three mixing angles [16,17,18]. 
We confine our analysis to CP-conserving case, neglecting all CP phases in the neutrino mixing matrix. 

 In the present paper we carry out a detailed numerical analysis of these two aprroaches for a consistency check 
on numerical accuracy, and 
find out the stability criteria of the main three neutrino mass models[19]. We give  all the zeroth-order as well as  full textures of the left-handed 
neutrino mass matrices obtained from seesaw mechanism in Appendix (A,B), and use these expressions for checking the stability criteria.
In addition, we further study the validity of some existing conjectures based on radiative corrections. The effect of running vacuum expectation value 
to the evolution of neutrino masses, is further examined in both approaches.  
The paper is organised as follows. In section 2, we briefly outline the main points of the two approaches on renornalisation group analysis. 
The numerical analysis and main results are presented in section 3. Section 4 concludes with a summary and discussion. 


\section{Renormalisation group analysis of  neutrino masses and mixings}
We present a very brief review on  the two main approaches of taking quantum radiative corrections of neutrino masses and mixings in MSSM.  
Our main motivation is to have a numerical consistency check on the results of these two approaches, and apply again to check the validity on some existing 
conjectures related to radiative corrections. 

 \subsection{Method A: Evolution of neutrino mass matrix}
In this approach  the quantum radiative corrections  are  taken on all the elements of the neutrino mass matrix $m_{LL}$  in the basis
 where charged lepton mass matrix is diagonal. The diagonalisation of the neutrino mass matrix at any  particular 
energy scale leads to the physical neutrino mass eigenvalues  as well as three  mixing angles.  The neutrino mass matrix
 $m_{LL}(t)$ which is generally obtained from seesaw mechanism, is expressible in term of 
$K(t)$, the coefficient of the dimension five neutrino mass operator in a scale-dependent manner,  $t=\ln (\mu/1GeV)$,
\begin{equation}
m_{LL}(t)=v_{u}^{2}K(t)
\end{equation}
where the vacuum expectation value (vev) is $v_{u}=v_{0}\sin\beta$, and $v_{0}=174$GeV in MSSM. The evolution equation
of the coefficient  $K(t)$ in the basis where the charged lepton mass matrix is diagonal, is given by[10]
\begin{equation}
\frac{d}{dt}K(t)=-\frac{1}{16\pi^{2}}\times[\frac{6}{5}g_{1}^2+6g_{2}^2-6h_{t}^2-h_{\tau}^{2}\delta_{i3}-h_{\tau}^{2}\delta_{3j}]
\end{equation}
 We  replace $K(t)$ by $m_{LL}(t)$ in the above equation where we assume that vev is scale-independent. Upon integration from high $(B-L)$ breaking scale 
$t_{R}(=\ln (M_{R}/1GeV)$ to top-quark mass scale $t_{0}(=\ln (m_{t}/1GeV))$ where $t_{0}\leq t\leq t_{R}$, we get 
\begin{equation}
m_{LL}(t_{0})=\left(\begin{array}{ccc}
m_{11}(t_{R}) & m_{12}(t_{R}) & m_{13}(t_{R})e^{-I_{\tau}(t_{0})}\\
m_{21}(t_{R}) & m_{22}(t_{R}) & m_{23}(t_{R})e^{-I_{\tau}(t_{0})}\\
 m_{31}(t_{R})e^{-I_{\tau}(t_{0})} & m_{32}(t_{R})e^{-I_{\tau}(t_{0})} & m_{33}(t_{R})e^{-2I_{\tau}(t_{0})}\\
\end{array}\right)R_{0},
\end{equation}
Here the overall factor $R_{0}$ which does not affect the mixing angles, is given by the expression,
\begin{equation}
R_{0}=exp[(6/5)I_{g1}(t_{0}) + 6I_{g2}(t_{0}) - 6I_{t}(t_{0})]
\end{equation} 
which strongly  depends on top-quark Yukawa coupling and gives very large contribution for small $\tan\beta$ value (=1.60). 
The integrals in the above expressions are defined as[7,10] 
\begin{equation}
I_{gi}(t_{0})=\frac{1}{16\pi^2}\int_{t_{0}}^{t_{R}}g_{i}^{2}(t)dt
\end{equation}
and,
\begin{equation}
I_{f}(t_{0})=\frac{1}{16\pi^2}\int_{t_{0}}^{t_{R}}h_{f}^{2}(t)dt
\end{equation}
where $i=1,2,3$ and $f=t,b,\tau$ respectively. The numerical values of these integrals at different energy scales, can be calculated 
from the running of the RGEs for three gauge couplings $(g_{1},g_{2},g_{3})$ and the third family Yukawa couplings $(h_{t},h_{b},h_{\tau})$ in  MSSM.
The mass eigenvalues and the MNS mixing matrix [20] are estimated through diagonalisation of $m_{LL}$ at every point in the energy scale,
\begin{equation}
m_{LL}^{diag}=Diag(m_{1},m_{2},m_{3})=V_{\nu L}m_{LL}V_{\nu L}^{T}
\end{equation}
and $V_{MNS}=V^{\dag}_{\nu L}$ which is identified with $U_{fi}$ in the neutrino oscillation relation
\begin{equation}
|\nu_{f}>=U_{fi}|\nu_{i}>
\end{equation}
where $f=\tau,\mu,e$ and $i=1,2,3$. The MNS mixing matrix[20]
\begin{equation}
U_{MNS}=\left(\begin{array}{ccc}
U_{e1} & U_{e2} & U_{e3} \\
U_{\mu 1} & U_{\mu 2} & U_{\mu 3} \\
U_{\tau 1} & U_{\tau 2} & U_{\tau 3}
\end{array}\right)
\end{equation}
 is usually  parametrised in term of three rotations (neglecting CP violating phases) by 
\begin{equation}
U_{MNS}=\left(\begin{array}{ccc}
c_{13}c_{12}  & c_{13}s_{12}  & s_{13} \\
-c_{23}s_{12}-c_{12}s_{13}s_{23}  & c_{12}c_{23}-s_{12}s_{13}s_{23} & c_{13}s_{23} \\
s_{12}s_{23}-c_{12}s_{13}c_{23} & -c_{12}s_{23}-c_{23}s_{13}s_{12} & c_{13}c_{23}
\end{array}\right)
\end{equation}
where $s_{ij}=\sin{\theta_{ij}}$ and $c_{ij}=\cos{\theta_{ij}}$  respectively. The unitarity conditions are also satisfied
$$U_{e1}^2+U_{e2}^2+U_{e3}^2=1=U_{e3}^2+U_{\mu 3}^2+U_{\tau 3}^2$$
Hence $\tan\theta_{12}=|U_{e2}|/|U_{e1}|$ and $\tan\theta_{23}=|U_{\mu 3}|/|U_{\tau 3}|$ and $\sin\theta_{13}=|U_{e3}|$.
The solar LMA MSW solution  favours the 'light-side' $\tan\theta_{12}<1$ of the data [21,22,23] for the usual sign convention $|m_2|>|m_1|$.
It is also possible to express mixing angles in terms of $\sin\theta_{ij}$ directly[24].

\subsection{Method B: Evolution of neutrino mass eigenvalues and mixing angles}
Here we follow  the main expressions  from Ref.[16] for the evolution of three neutrino mass eigenvalues 
and mixing angles{\footnote{In Ref.[18] the RGEs for the neutrino masses, mixing angles and CP phases are also derived,
 whereas in Ref.[17] a thorough discussion on fine-tuning between initial conditions and radiative corrections is given for quasi-degenerate 
neutrino masses for two generations.}.
 In this approach  the RGEs for the  eigenvalues of coefficient $K(t)$ can be  expressible as 
\begin{equation}
\frac{d}{dt}K_{i}=\frac{1}{16\pi^2}\sum_{f=e,\mu,\tau}[(-\frac{6}{5}g_{1}^{2}-6g_{2}^2+6Tr(h_{u}^2)+2h_{f}^{2}U_{fi}^{2}]K_{i}
\end{equation}
Neglecting $h_{\mu}^2$ and $h_{e}^2$ compared to $h_{\tau}^2$, and taking the scale-independent vev as before, we have the complete RGEs
 for three neutrino mass eigenvalues[16], 
\begin{equation}
\frac{d}{dt}m_{i}=\frac{1}{16\pi^2}[(-\frac{6}{5}g_{1}^{2}-6g_{2}^2+6h_{t}^2)+2h_{\tau}^{2}U_{\tau i}^{2}]m_{i}
\end{equation}
One can have an approximate analytical solution of the above equation by neglecting the small effect due to the change 
of $U_{\tau i}^2$ in the integration range, as[15]
\begin{equation}
m_{i}(t_{0})=m_{i}(t_R)exp(\frac {6}{5} I_{g1}+6I_{g2}-6I_{t})exp(-2U_{\tau i}^2 I_{\tau})
\end{equation} 
These equations  lead to the equations derived in JM conjecture[11] and have interesting consequences. However it should be 
emphasised that these equations are valid when the mixing angles are only static.
The corresponding evolution equations for the MNS matrix elements $U_{fi}$ are given by [16]
\begin{equation}
\frac{d U_{fi}}{dt}=-\frac{1}{16\pi^2}\sum_{k\neq i}\frac{m_{k}+m_{i}}{m_{k}-m_{i}} U_{fk}(U^TH^2_eU)_{ki}
\end{equation}
where  $f=e,\mu,\tau$ and $i=1,2,3$ respectively. Neglecting $h_{\mu}^2$ and $h_{e}^2$ as before, one can  simplify the terms,
$$(U^TH_e^2U)_{13}\simeq h_{\tau}^2(U^T_{1\tau}U_{\tau 3})$$
$$(U^TH_e^2U)_{23}\simeq h_{\tau}^2(U^T_{2\tau}U_{\tau 3})$$ 
 $$(U^TH_e^2U)_{12}\simeq h_{\tau}^2(U^T_{1\tau}U_{\tau 2})$$
 Denoting $A_{ki}=\frac{m_{k}+m_{i}}{m_{k}-m_{i}}$, we can write the RGEs for all the elements of MNS matrix. For example,
 we give here only three of them relevant to our requirement,
\begin{equation}
\frac{d U_{e2}}{dt}=-\frac{1}{16\pi^2}[U_{\tau 2}h_{\tau}^2(A_{32}U_{e3}U^{\dag}_{3\tau}+A_{12}U_{e1}U^{\dag}_{1\tau})]
\end{equation}
\begin{equation}
\frac{d U_{e3}}{dt}=-\frac{1}{16\pi^2}[U_{\tau 3}h_{\tau}^2(A_{13}U_{e1}U^{\dag}_{1\tau}+A_{23}U_{e2}U^{\dag}_{2\tau})]
\end{equation}
\begin{equation}
\frac{d U_{\mu 3}}{dt}=-\frac{1}{16\pi^2}[U_{\tau 3}h_{\tau}^2(A_{13}U_{\mu 1}U^{\dag}_{1\tau}+A_{23}U_{\mu 2}U^{\dag}_{2\tau})]
\end{equation}
Using the MNS parametrisation in eq.(10), the above three expressions(15) - (17) simplify to [16] 
\begin{equation}
\frac{d s_{12}}{dt}=\frac{1}{16\pi^2}h_{\tau}^2 c_{12}[c_{23}s_{13}s_{12}U_{\tau 1}A_{31}-c_{23}s_{13}c_{13}U_{\tau 2}A_{32}
+U_{\tau 1}U_{\tau 2} A_{21}]
\end{equation}

\begin{equation}
\frac{d s_{13}}{dt}=\frac{1}{16\pi^2}h_{\tau}^2 c_{23}c_{13}^2[c_{12}U_{\tau 1}A_{31}+s_{12}U_{\tau 2}A_{32}]
\end{equation}

\begin{equation}
\frac{d s_{23}}{dt}=\frac{1}{16\pi^2}h_{\tau}^2 c_{23}^2[-s_{12}U_{\tau 1}A_{31}+c_{12}U_{\tau 2}A_{32}]
\end{equation}
 \subsection{Effect of scale-dependent vev}
In this section we modify the results of Ref.[16] by 
 considering  the running of the vev $v_{u}(t)$ through    the neutrino mass formula    $m_{i}(t)=v_{u}^2(t) K_{i}(t)$. This gives [15] 
\begin{equation}
\frac{d (ln m_{i})}{dt}=\frac{d (ln K_{i})}{dt}+2\frac{d (ln v_{u})}{dt}
\end{equation}
where the RGE for $v_u$ in MSSM is given by 
\begin{equation}
\frac{d}{dt}v_{u}=\frac{1}{16\pi^2}[\frac{3}{20}g_{1}^{2}+\frac{3}{4}g_{2}^2-3h_{t}^2]v_{u}
\end{equation}
The complete RGEs for neutrino mass eigenvalues are now  given by [15]
\begin{equation}
\frac{d}{dt}m_{i}=\frac{1}{16\pi^2}[(-\frac{9}{10}g_{1}^{2}-\frac{9}{2}g_{2}^2)+2h_{\tau}^{2}U_{\tau i}^{2}]m_{i}
\end{equation} leading to the approximate solution, 
\begin{equation}
m_{i}(t_{0})=m_{i}(t_R)exp(\frac {9}{10} I_{g1}+\frac{9}{2} I_{g2})exp(-2U_{\tau i}^2 I_{\tau})
\end{equation}
Similarly, for Method A  with the inclusion of scale-dependent vev, $R_{0}$ in eq.(4) is now replaced by  
\begin{equation}
R_{0}=exp(\frac{9}{10}I_{g1}+\frac{9}{2}I_{g2})
\end{equation}
Now the top-quark dependent term has been cancelled out  and this  will certainly affect the overall magnitude of the neutrino masses,
but not the mixing angles. With this modification the magnitudes of neutrino masses tend to increase with the decrease of energy scale. 
The enhancement factor in the magnitude of neutrino masses through RGEs is calculated as 

\begin{equation}
f=exp(6I_{t}-\frac{3}{10}I_{g1}-\frac{3}{2}I_{g2})
\end{equation}
which gives a positive numerical value greater than one  even for large $\tan\beta$ values $(=55)$.

\section{Numerical analysis  and results}
For a complete numerical analysis of the RGEs for both Methods A and B presented in the previous section,
we follow  here two steps (a) bottom-up running[7]  and (b)  top-down running[10]. In the first step (a), the running 
of the RGEs for the third family Yukawa couplings $(h_{t},h_{b}, h_{\tau})$ and three gauge couplings $(g_{1},g_{2},g_{3})$ 
in MSSM , is carried out  from top-quark mass scale ($t_0$) at low energy end to high energy scale $M_{R}$ where $B-L$ symmetry breaks down[7,15].
In the present analysis we consider the high scale  $M_{R}=10^{13}$GeV and take the large $\tan\beta$ input value ( $\tan\beta=55$). 
For  simplicity of the calculation, we assume here  the supersymmetric breaking scale at the top-quark 
mass scale $t_0=\ln m_t$[7]. We adopt the  standard procedure to get the 
 values of gauge couplings at top-quark mass scale from the experimental CERN-LEP measurements 
at $M_{Z}$, using one-loop RGEs, assuming the existence of a 
one-light Higgs doublet and five quark flavours below $m_t$ scale[15]. Similarly, the  Yukawa couplings are 
also evaluated at top-quark mass scale using QCD-QED 
rescaling factors in the standard fashion[15]. 
We present here the  values of the Yukawa couplings and gauge couplings at two scales $t_0=5.159$ and $t_R=29.954$
 as follows:\\
\\

\begin{tabular}{lll}\hline
 $t_0=5.159$ & $t_{R}=29.954$ & $t_R=29.954$ \\  \hline

$h_t$= 1.000165276 & $h_t$= 0.849373996 & $I_t$=0.109299302 \\
$h_b$ = 0.866235097 & $h_b$= 0.641425729 & $I_b$=0.0759593919\\
$h_{\tau}$ = 0.555977506 & $h_{\tau}$=0.679428339 & $I_{\tau}$=0.0620016195\\
$g_1$ = 0.463751 & $g_1$=0.626455069  & $I_{g1}$= 0.0544101298\\
$g_2$ = 0.6513289 & $g_2$=0.708234191  & $I_{g2}$=0.0803824514\\
$g_3$ = 1.1891966 & $g_3$=0.784410238 & $I_{g3}$= 0.141634345\\ \hline
\end{tabular}
\\

The values of the integrals  are estimated between the two limits $(t_0, t_R)$.  In the second stage (b), the  running of
 three neutrino masses $(m_1, m_2, m_3)$  and mixing angles  $(s_{12}, s_{13}, s_{23})$ is carried out together with the 
running of the  Yukawa and gauge cuoplings,
 from the high scale $t_R$ to low scale $t_0$. In this case we use the values of Yukawa and gauge couplings evaluated earlier 
 at the scale $t_R$ from the first stage runnung of RGEs.
In principle one can evaluate neutrino masses and mixing angles at every point in the energy scale.  

We present the results of our numerical analysis in Tables 1-4.  
First we check the stability of the neutrino mass models under radiative corrections. Table 1(a) and (b) give the values of 
neutrino masses and mixing angles at high and low scales for three neutrino mass models -  hierarchical,
inverted hierarchical and degenerate models (ses Appendix- A,B). For a check on numerical consistency, we evaluate the quantities 
for both Methods A and B outlined in the previous section. Both methods 
nearly  give consistent results to about a factor of  $13\%$ discrepancy. Only the hierarchical model (Type III) and the inverted hierarchical model (IIB)
 with opposite CP parity, are found to be stable under RG analysis in MSSM. In fact,  both 
$\bigtriangleup m_{12}^2$ and $\bigtriangleup m_{23}^2$ are slightly reduced, whereas the mixing angles are slightly increased with 
the decrease in energy scale. But there is no 
substantial change in both parameters, which may  spoil the good predictions already achieved at the  high energy scale. 
In case of inverted hierarchical model  with same CP parity ( IIA),
 the solar angle $(s_{12})$ is not stable under  radiative corrections. Similarly,  in all the three degenerate models (I-A,B,C) 
both solar and atmospheric mass scales 
as well as solar and atmospheric angles, are not stable under radiative corrections.  

In Table 2 we analyse the MPR conjecture [12] which states  that quark and lepton mixing angles  are 
identical at high energy scale, and large solar and atmospheric neutrino mixing angles together with the small reactor 
angle, can be  understood purely as a result of RG evolution provided the three neutrino masses  are quasi-degenerate and have the same CP parity.
In the present numerical analysis in Method B, both Yukawa and gauge couplings are running together with neutrino masses and mixings. It appears 
that in MPR paper[12]  only neutrino masses and mixings are running.  We present here  three set of readings ( cases (i),(ii),(iii))
which give good radiative magnifications. However in case (iv) we use the same input values quoted in MPR paper but we observe less  magnification contrary to 
MPR paper[12]. This discrepancy may be due to  the  running of gauge and Yukawa couplings along with neutrino masses 
and mixings in the form of coupled equations in the present analysis.
In short, MPR conjecture[12] is well verified.  The analysis presented here in Table 2 is the results of Method B, but 
we find  that both methods (A and B)  give consistent results.  

In Table 3 we supply some new results connected to two more conjectures on  radiative corrections in MSSM.  JM conjecture [11]  specifies that 
 radiative corrections can generate the neutrino mass-squared difference required for the large mixing angle (LMA)
MSW solution to the solar neutrino problem if two of the three neutrino masses  are assumed to be exactly degenerate $(m,-m,m')$ at
 high energy scale,  and also  if $U_{e3}$ 
at high scale is non-zero. We have shown in Table 3 that for a limited range of non-zero values of  $m'$, it is also possible to generate LMA MSW solution even 
if $U_{e3}$ 
is zero at high scale. If both $m'$  and $U_{e3}$ are zeros at high scale, then it is not possible to generate LMA MSW solution.
 Case (ii) in Table 3 is interesting in the 
sense that it gives exception to JM conjecture by taking $U_{e3}=0$ and $m'=0.07eV$. At low energy we get $\bigtriangleup m_{12}^2=5.23\times 10^{-5}eV^2$
and $\sin\theta_{12}=0.7042421$ which is high but lies in the light side of the data ($\tan\theta_{12}<1$) [21,22,23].
 Fig.1 shows the evolution of the three mixing angles,  and 
the CHOOZ angle $\sin\theta_{13}$ is generated through radiative corrections. Fig.2 presents the evolution of the three neutrino masses,
 and the solar mass scale generated through 
radiative corrections at low scale, is then demonstrated in Fig.3. It is still necessary to tone down the solar mixing angle through further fine tuning. 
As an  example, using $s_{12}=0.7$ at high scale, one can get $s_{12}=0.697$ at lower scale.   
We also discuss MST conjecture [14] which states  that starting from bimaximal mixings at high scale, radiative corrections due to the $\tau$-Yukawa 
coupling leads to solar angle towards the dark side at low scale,($\tan\theta_{12}>1$). The results presented in Table 3 also show that MST conjecture [14]
is not always valid for the cases (i)-(iv) discussed here. \\

Finally, as expected, the effect of running vev in neutrino mass formula leads to the  increase of neutrino masses with the decrease of energy scale[15].
In Table 4  we  calculate the neutrino masses and mixings with (I) or without (II) the  effect of running the vev,  at low energy scale. There 
is  a factor of about  $1.91$ higher in case of  low-energy neutrino mass values obtained with the  running effect of  vev. 
Figs. 4 and 5 show  the evolution of neutrino masses for the above two cases (II) and (I) respectively.

\pagebreak
Table 1(a): Running of neutrino masses and mixing angles from high scale $M_R=10^{13}$GeV to top-quark mass scale $m_t=175$GeV
in MSSM for hierarchical  (Type III) and inverted hierarchical  (II) models ($m_{LL}$ collected from Appendix B). 
 Methods A and B are explained in the text.

\begin{tabular}{lllll}\hline
Type  & Item &  $\mu=M_{R}$  & (A):  $\mu=m_{t}$ & (B): $\mu=m_{t}$ \\ \hline
III&$m_{1}$     & 0.00336eV        & 0.0029358eV          &   0.002502eV           \\
& $m_{2}$     & 0.007357eV       & 0.0063875eV          &   0.00534eV            \\
& $m_{3}$     & 0.057013eV       & 0.0479200eV          &   0.040322eV           \\
& $s_{23}$    & 0.65630          & 0.6632               &   0.68441            \\
& $s_{13}$    & 0.07358          & 0.08141              &   0.07911            \\
& $s_{12}$    & 0.5838           & 0.58807              &   0.608521           \\
& $\bigtriangleup m_{12}^{2}$  & $4.28\times10^{-5}eV^2$  &  $3.22\times10^{-5}eV^2$  & $2.23\times10^{-5}eV^2$ \\
& $\bigtriangleup m_{23}^{2}$  & $3.20\times10^{-3}eV^2$  &  $2.20\times10^{-3}eV^2$  & $1.60\times10^{-3}eV^2$ \\ \hline
IIB & $m_{1}$     & -0.070445eV        &-0.0613552eV          &   -0.05171696eV           \\
& $m_{2}$     & 0.070977eV       & 0.0618069eV          &   0.0520974eV            \\
& $m_{3}$     & 0.0005324eV       & 0.0004482eV          &   0.000378eV           \\
& $s_{23}$    & 0.7071          & 0.68486               &   0.685182            \\
& $s_{13}$    & 0.0          & 0.00045              &   0.000445            \\
& $s_{12}$    & 0.7057745           & 0.70581              &   0.7057745           \\
& $\bigtriangleup m_{12}^{2}$  & $7.52\times10^{-5}eV^2$  &  $5.56\times10^{-5}eV^2$  & $4.22\times10^{-5}eV^2$ \\
& $\bigtriangleup m_{23}^{2}$  & $4.96\times10^{-3}eV^2$  &  $3.76\times10^{-3}eV^2$  & $2.67\times10^{-3}eV^2$ \\ \hline
IIA & $m_{1}$     & 0.0497257eV        & 0.042245eV          &   0.0356237eV           \\
& $m_{2}$     & 0.0500693eV       & 0.0446987eV          &   0.037661eV            \\
& $m_{3}$     & 0.000005eV       & 0.000004eV          &   0.000003eV           \\
& $s_{23}$    & 0.707107          & 0.68486               &   0.68518            \\
& $s_{13}$    & 0.0          & 0.0             &   0.0            \\
& $s_{12}$    & 0.465           & 0.99874             &   0.998697           \\
& $\bigtriangleup m_{12}^{2}$  & $3.43\times10^{-5}eV^2$  &  $21.13\times10^{-5}eV^2$  & $14.93\times10^{-5}eV^2$ \\
& $\bigtriangleup m_{23}^{2}$  & $2.47\times10^{-3}eV^2$  &  $2.00\times10^{-3}eV^2$  & $1.42\times10^{-3}eV^2$ \\ \hline
\end{tabular}
\pagebreak

Table 1(b): Running of neutrino masses and mixing angles from high scale $M_R=10^{13}$GeV to top-quark mass scale $m_t=175$GeV
in MSSM for degenerate  models ($m_{LL}$ collected from Appendix B). Methods A and B are explained in the text.

\begin{tabular}{lllll}\hline
Type  & Item &  $\mu=M_{R}$  & (A):  $\mu=m_{t}$ & (B): $\mu=m_{t}$ \\ \hline
IA & $m_{1}$     & 0.396484eV        & 0.326001eV          &   0.275101eV           \\
&$m_{2}$     & -0.396532eV       & -0.345072eV          &  -0.290815eV            \\
& $m_{3}$     & 0.4eV       & 0.396941eV          &   0.3007399eV           \\
& $s_{23}$    & 0.707107          & 0.99885               &   0.998810            \\
& $s_{13}$    & 0.0          & 0.556             &   0.558            \\
& $s_{12}$    & 0.70931           & 0.853267             &   0.85202           \\
& $\bigtriangleup m_{12}^{2}$  & $3.81\times10^{-5}eV^2$  &  $12.80\times10^{-3}eV^2$  & $8.89\times10^{-3}eV^2$ \\
& $\bigtriangleup m_{23}^{2}$  & $2.76\times10^{-3}eV^2$  &  $3.85\times10^{-2}eV^2$  & $5.87\times10^{-3}eV^2$ \\ \hline
IB& $m_{1}$     & 0.396841eV        & 0.315833eV          &   0.266586eV           \\
& $m_{2}$     & 0.396891eV       & 0.35615eV          &  0.3000754eV            \\
& $m_{3}$     & 0.4eV       & 0.357618eV          &   0.301312596eV           \\
& $s_{23}$    & 0.707107          & 0.99950               &   0.99949            \\
& $s_{13}$    & 0.0          & 0.00854             &   0.011254            \\
& $s_{12}$    & 0.459701           & 0.99999             &   0.99999           \\
& $\bigtriangleup m_{12}^{2}$  & $3.97\times10^{-5}eV^2$  &  $27.09\times10^{-3}eV^2$  & $18.98\times10^{-3}eV^2$ \\
& $\bigtriangleup m_{23}^{2}$  & $2.48\times10^{-3}eV^2$  &  $1.05\times10^{-3}eV^2$  & $0.74\times10^{-3}eV^2$ \\ \hline
IC & $m_{1}$     & 0.396841eV        & 0.334761eV          &   0.2823071eV           \\
& $m_{2}$     & 0.396891eV       & 0.35615eV          &  0.3000757eV            \\
& $m_{3}$     & -0.4eV       & -0.337397eV          &   -0.284533eV           \\
& $s_{23}$    & 0.707107          & 0.7071               &   0.70722            \\
& $s_{13}$    & 0.0          & 0.0             &   0.0            \\
& $s_{12}$    & 0.4597           & 1             &   0.999999           \\
& $\bigtriangleup m_{12}^{2}$  & $3.97\times10^{-5}eV^2$  &  $14.78\times10^{-3}eV^2$  & $10.35\times10^{-3}eV^2$ \\
& $\bigtriangleup m_{23}^{2}$  & $2.48\times10^{-3}eV^2$  &  $13.01\times10^{-3}eV^2$  & $9.09\times10^{-3}eV^2$ \\ \hline
\end{tabular}
\pagebreak
Table 2: Analysis on MPR conjecture[12]  related to radiative magnification on solar and atmospheric mixing angles at low scale.
The parameters $(m^0_{1,2,3}, s^0_{23,13,12})$ are defined at high scale $M_R=10^{13}$GeV and others are defined at low scale $m_t=175$GeV.
Cases (i)-(iv) include different sets of arbitrary input parameters.

\begin{tabular}{llll} \\ \hline
Case (i) & Case (ii) & Case  (iii) & Case (iv) \\  \hline 
$m_{1}^0=0.5288446$eV   & 0.5288446eV  & 0.5288446eV &    0.3682eV \\
$m_{2}^0=0.5309554$eV   & 0.5309554eV &  0.5309554eV &   0.370eV \\
$m_{3}^0=0.6$eV         & 0.6 eV      &  0.6eV &       0.421eV  \\
$s_{23}^0=0.0311$        & 0.03     &  0.031  &   0.038 \\
$s_{13}^0=0.005$        & 0.0      &   0.004  &   0.0025 \\
$s_{12}^0=0.22$         & 0.2      &   0.215  &   0.22 \\ \hline
$m_{1}=0.399836034$eV   & 0.3997283eV  &  0.39677897eV &  0.2783897eV \\
$m_{2}=0.399989337$eV   & 0.4001846eV &   0.396988094eV &  0.2791149eV \\
$m_{3}=0.403035134$eV    & 0.40294793eV & 0.39996222eV & 0.28240252eV  \\
$s_{23}=0.709379$        & 0.71158975 & 0.71025485    & 0.467712 \\
$s_{13}=0.135225$        & 0.071834   & 0.12237   & 0.05389 \\
$s_{12}=0.656099$         & 0.6637173   & 0.670588    & 0.32458 \\ 
$\bigtriangleup m_{12}^2=12.26\times10^{-5}eV^2$ & $36.50\times10^{-5}eV^2$ &  $ 16.60\times 10^{-5}eV^2$ & $40.43\times 10^{-5}eV^2$ \\
$\bigtriangleup m_{23}^2=2.45\times10^{-3}eV^2$ & $2.22\times10^{-3}eV^2$ & $2.37\times 10^{-3}eV^2$ & $1.85\times 10^{-3}eV^2$ \\ \hline
\end{tabular}
\\
\pagebreak

Table 3: Analysis on JM  conjecture[11]  related to radiative magnification on solar mass scale  at low scale.
The parameters $(m^0_{1,2,3}, s^0_{23,13,12})$ are defined at high scale $M_R=10^{13}$GeV and others are defined at low scale $m_t=175$GeV.
Cases (i)-(iv) include different sets of arbitrary input parameters.

\begin{tabular}{llll} \\ \hline
Case (i) & Case (ii) &Case  (iii) & Case (iv) \\  \hline 
$m_{1}^0=-0.095$eV                        & -0.095eV     & -0.08 eV   & 0.08eV\\
$m_{2}^0=0.095$eV                       & 0.095eV     & -0.08eV  & -0.08eV\\
$m_{3}^0=0.0$eV                          & 0.07eV      & 0.04eV   & 0.0eV\\
$s_{23}^0=0.707107$                      & 0.707107     & 0.707107 & 0.707107 \\
$s_{13}^0=0.1$                           & 0.0          & 0.1  & 0.0 \\
$s_{12}^0=0.707107$                      & 0.707107     & 0.707107  & 0.707107 \\ \hline
$m_{1}=-0.0693038$eV                     & -0.06968284eV  & -0.0583183eV  & -0.058726eV\\
$m_{2}=0.07013016$eV                     & 0.070057414eV  &  0.0591391eV  & 0.058726eV\\
$m_{3}=0.0$eV                            & 0.049511038eV   & 0.0283987eV   & 0.0eV\\
$s_{23}=0.6852$                          & 0.638701      & 0.6745     & 0.6852\\
$s_{13}=0.0969$                          & 0.087023      & 0.1329     & 0.0\\
$s_{12}=0.705008$                        & 0.7042421     & 0.70341    & 0.707107\\ 
$\bigtriangleup m_{12}^2=11.52\times10^{-5}eV^2$ & $5.23\times10^{-5}eV^2$ & $9.64\times 10^{-5}eV^2$  & $0.0eV^2$ \\
$\bigtriangleup m_{23}^2=4.92\times10^{-3}eV^2$ & $2.46\times10^{-3}eV^2$ & $2.69\times 10^{-3}eV^2$  & $3.45\times 10^{-3}eV^2$\\ \hline
\end{tabular}

\pagebreak
Table 4: Running of neutrino mass at low scale with (I)  and without (II)  the effect of scale- dependent vev $v_u^2$. 

\begin{tabular}{llll}\hline
 Item &  $\mu=M_{R}$  &   $\mu=m_{t}$(II) &  $\mu=m_{t}$(I) \\ \hline
 $m_{1}$     & 0.528844eV        & 0.399836034eV          &   0.762722015eV           \\
 $m_{2}$     & 0.5309554eV       & 0.399989337eV          &   0.763014257eV            \\
 $m_{3}$     & 0.6eV       & 0.403035134eV          &   0.768824458eV           \\
 $s_{23}$    & 0.0311          & 0.709379              &   0.709379           \\
 $s_{13}$    & 0.005          & 0.135225            &   0.135225        \\
 $s_{12}$    & 0.22           & 0.656099             &   0.656099           \\
 $\bigtriangleup m_{12}^{2}$  & $2.24\times10^{-3}eV^2$  &  $12.26\times10^{-5}eV^2$  & $44.588\times10^{-5}eV^2$ \\
 $\bigtriangleup m_{23}^{2}$  & $7.81\times10^{-2}eV^2$  &  $2.45\times10^{-3}eV^2$  & $8.90\times10^{-3}eV^2$ \\ \hline
\end{tabular}
\section{Summary and Discussion}
We summarise the main points in this work. First we briefly review the main points of the formalism based on  
two  approaches on the evolution of RGEs of neutrino masses and mixings. The first one (A) deals with the running of the whole 
neutrino mass matrix from which one can extract mass eigenvalues and mixings at any particular energy scale, whereas in the second approach (B)
the three  neutrino mass eigenvalues and  the three mixing angles are running directly. Detailed numerical analysis shows that both 
approaches agree up to a discrepancy of a factor of $13\%$ in mass eigenvalues. The predictions on mixing angles are almost consistent
 in these two approaches. Using both approaches we show that  hierarchical model(III) and inverted hierarchical model with opposite 
CP parity (IIB) are   stable under radiative corrections. The evolution of $\sin\theta_{12}$ is very fast in the inverted hierarchical model 
with same CP parity (IIA), 
and hence the model is not stable. We also verify the MPR conjecture [12] in which radiative magnification of solar and atmospheric mixings 
are possible in case of nearly degenerate model with same CP parity. We find  that runnings of masses and mixings with  
and without the running of gauge and Yukawa couplings, give reasonably different magnifications. However such 
 radiative amplification  generally involves  a delicate fine-tuning of  the initial conditions which are to some extent unnatural[17].
 However such problems are not there in  the theory of neutrino masses derived from Kahler potential in supersymmetric model, 
and neutrino mixings angles can easily be driven to large values at low energy as they approach infrared pseudo-fixed points at large mixing [17].  
 
 We also study JM conjecture [11] which 
specifies the radiative generation of solar scale in exactly two-fold degenerate model having opposite CP parity $(m,-m,m_3)$ and non-zero values of  $U_{e3}$.
 Our numerical analysis shows that the same radiative generation of solar mass scale is also  possible with the conditions $U_{e3}=0.0$ and non-zero
 value of  $m_{3}$. We also discuss 
the MST conjecture[14]  which 
states that starting from bimaximal mixings at high scale, radiative corrections lead to the solar angle towards the dark side of the data at low energy scale. 
We show that MST conjecture is not always valid.
We suggest further generalisation of  JM  conjecture and this will be reported in subsequent communication[25].
  Finally, the effect of running the vev in neutrino mass renormalisation is discussed and it
 is observed that neutrino mass increases with the decrease of energy scale when we include the running  of vev. This gives magnification of a factor of  $1.91$
in neutrino masses at low energy scale, compared to the values calculated without the running vev effect. 
Numerical  analysis in   three generations with arbitrary CP violating phases is of great interest[18] 
and there has been  a possibility that within a restricted range of  the physical parameters including phases, the degenerate models are found to be stable
 under radiative corrections in MSSM [26]. As  emphasised before, oversimplifications of analytic expressions and a departure 
from the simultaneouse running of Yukawa and gauge couplings along with neutrino masses and mixings, may have a danger of getting misleading conclusions. 


\section*{Acknowledgements}
N.N.S. thanks the High Energy Physics Group, International Centre for Theoretical Physics, Trieste, Italy, 
for kind hospitality during the course of the work.

\pagebreak
\section*{Appendix A}

The  zeroth-order left-handed Majorana neutrino mass 
matrices with texture zeros, $m_{LL}$, corresponding to three models of neutrinos, viz., 
degenerate (Type [I]), inverted hierarchical (Type [II]) and normal hierarchical (Type [III]). 
These mass matrices are competible with the  LMA MSW solution as well as maximal atmospheric mixings[19].

\begin{center}
\begin{tabular}{ccc}\hline
Type  & $m_{LL}$ & $m_{LL}^{diag}$\\ \hline \\
\ [IA]    &${ \left(\begin{array}{ccc}
  0 & \frac{1}{\sqrt{2}} & \frac{1}{\sqrt{2}}\\ 
 \frac{1}{\sqrt{2}} & \frac{1}{2} & -\frac{1}{2}\\
 \frac{1}{\sqrt{2}} & -\frac{1}{2} & \frac{1}{2} 
\end{array}\right)}m_{0}$ & $Diag(1,-1,1)m_{0}$\\

\\
\ [IB]    &${ \left(\begin{array}{ccc}
  1 & 0 & 0\\ 
 0 & 1 & 0\\
 0 & 0 & 1 
\end{array}\right)}m_{0}$ & $Diag(1,1,1)m_{0}$\\
\\

\ [IC]    &${ \left(\begin{array}{ccc}
  1 & 0 & 0\\ 
 0 & 0 & 1\\
 0 & 1 & 0 
\end{array}\right)}m_{0}$ & $Diag(1,1,-1)m_{0}$\\ \hline
\\
\ [IIA]    &${ \left(\begin{array}{ccc}
  1 & 0 & 0\\ 
 0 & \frac{1}{2} & \frac{1}{2}\\
 0 & \frac{1}{2} & \frac{1}{2} 
\end{array}\right)}m_{0}$ & $Diag(1,1,0)m_{0}$\\
\\

\ [IIB]    &${ \left(\begin{array}{ccc}
  0 & 1 & 1\\ 
 1 & 0 & 0\\
 1 & 0 & 0 
\end{array}\right)}m_{0}$ & $Diag(1,-1,0)m_{0}$\\ \hline
\\

\ [III]    &${ \left(\begin{array}{ccc}
  0 & 0 & 0\\ 
 0 & \frac{1}{2} & -\frac{1}{2}\\
 0 & -\frac{1}{2} & \frac{1}{2} 
\end{array}\right)}m_{0}$ & $Diag(0,0,1)m_{0}$  \\ 

\\ \hline
\end{tabular}

\end{center}

\pagebreak
\section*{Appendix B}
The left-handed Majorana mass matrix $m_{LL}$  for three different models of neutrinos presented 
in Appendix A. These results are collected from Ref.[19].
\\

\underline{Degenerate model( Type [IA])}:\\
\\
$$m_{LL}= \left(\begin{array}{ccc}
  (-2\delta_{1}+2\delta_{2}) & ({\frac{1}{\sqrt{2}}}-\delta_{1}) & ({\frac{1}{\sqrt{2}}}-\delta_{1})\\ 
({ \frac{1}{\sqrt{2}}}-\delta_{1}) & ({\frac{1}{2}}+\delta_{2}) &({ -\frac{1}{2}}+\delta_{2})\\
 ({\frac{1}{\sqrt{2}}}-\delta_{1}) &({ -\frac{1}{2}}+\delta_{2}) &({ \frac{1}{2}}+\delta_{2}) 
\end{array}\right)0.4$$
\\
\underline{Degenerate model (Type [IB])}:\\
\\
$$m_{LL}= \left(\begin{array}{ccc}
  (1-2\delta_{1}-2\delta_{2}) & -\delta_{1} & -\delta_{1}\\ 
    -\delta_{1} & (1-\delta_{2}) & -\delta_{2}\\
 -\delta_{1} & -\delta_{2} & (1-\delta_{2}) 
\end{array}\right)0.4$$
\\
\underline{Degenerate model (Type [IC])}:\\
\\
$$m_{LL}= \left(\begin{array}{ccc}
  (1-2\delta_{1}-2\delta_{2}) & -\delta_{1} & -\delta_{1}\\ 
    -\delta_{1} & -\delta_{2} & (1-\delta_{2})\\
 -\delta_{1} & (1-\delta_{2}) & -\delta_{2} 
\end{array}\right)0.4$$
\\
\underline{Invereted hierarchical model(Type [IIA])}:\\
$$m_{LL}= \left(\begin{array}{ccc}
  (1-2\epsilon) & -\epsilon & -\epsilon\\ 
 -\epsilon & \frac{1}{2} & (\frac{1}{2}-\eta)\\
 -\epsilon & (\frac{1}{2}-\eta) & \frac{1}{2} 
\end{array}\right)0.05$$
\\
\underline{Inverted hierarchical model(Type [IIB])}:\\
$$m_{LL}= \left(\begin{array}{ccc}
  0 & 1 & 1\\ 
 1 & \lambda^3 & 0\\
 1 & 0 & \lambda^3
\end{array}\right)0.05$$
\\
\underline{Hierarchical model (Type [III]}:\\
$$m_{LL}=\left(\begin{array}{ccc}
-\lambda^4 & \lambda & \lambda^3 \\
\lambda & 1-\lambda & -1 \\
\lambda^3 & -1 & 1-\lambda^3
\end{array}\right)0.03$$
\\

The values of the parameters used are: Type IA:  $\delta_{1}=0.0061875$, $\delta_{2}=0.0030625$, \ \
Type [IB] and [IC]: $\delta_{1}=3.6\times 10^{-5}$, $\delta_{2}=3.9\times 10^{-3}$, \ \
Type [IIA]: $\eta=0.0001$, $\epsilon=0.002$, and Type [IIB] and [III]: $\lambda=0.22$. All neutrino masses are in eV.

\pagebreak

\pagebreak
\vbox{
\noindent
\hfil
\vbox{
\epsfxsize=10cm
\epsffile [130 380 510 735] {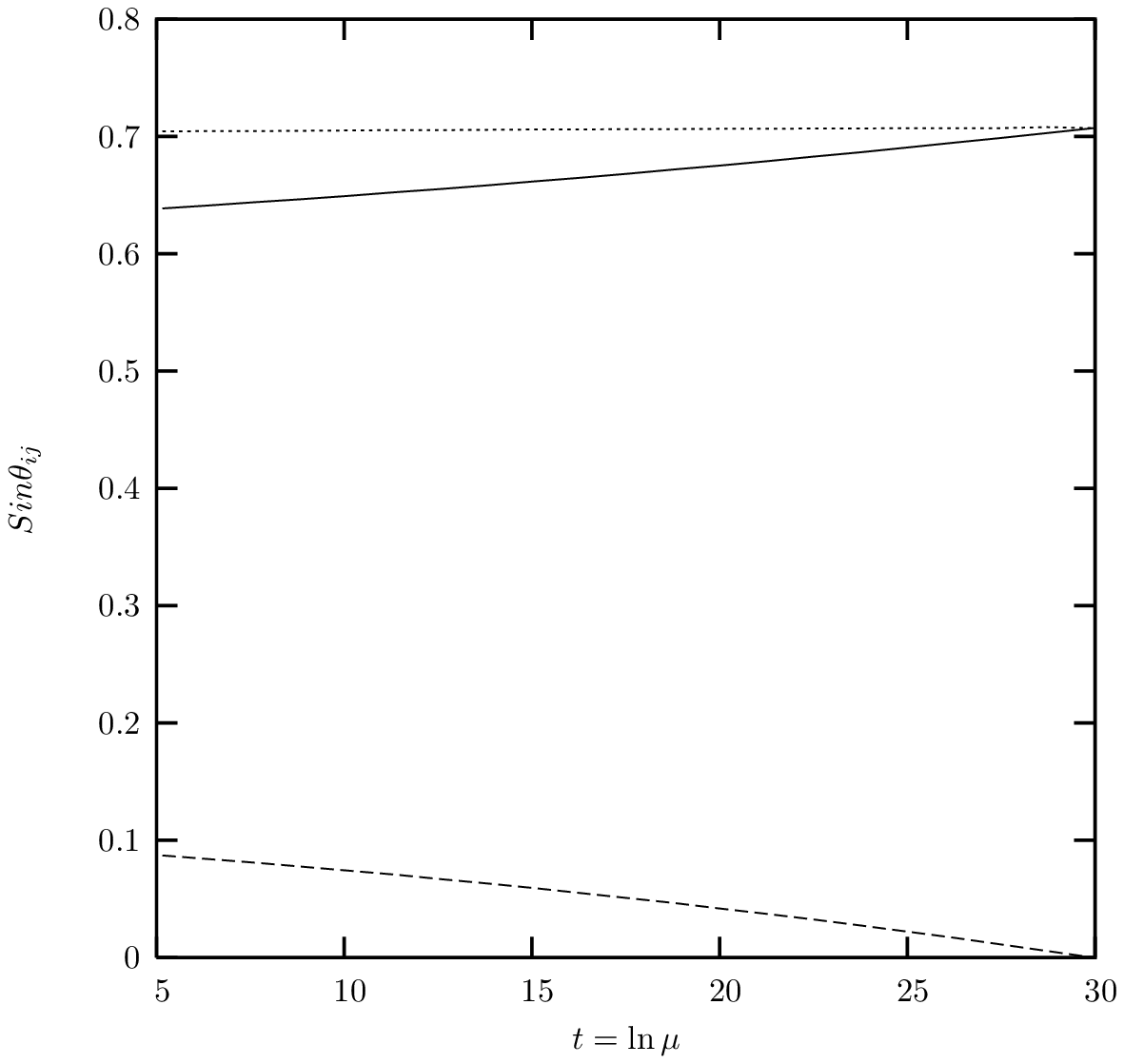}}

{\narrower\narrower\footnotesize\noindent
{Fig.1}
Evolution of the three mixing angles with energy scale  in JM conjecture[11]. $\sin\theta_{23}$, $\sin\theta_{13}$ and 
$\sin\theta_{12}$ are represented by solid line, dashed-line and dotted-line respectively.
\par}}

\vbox{
\noindent
\hfil
\vbox{
\epsfxsize=10cm
\epsffile [130 380 510 735] {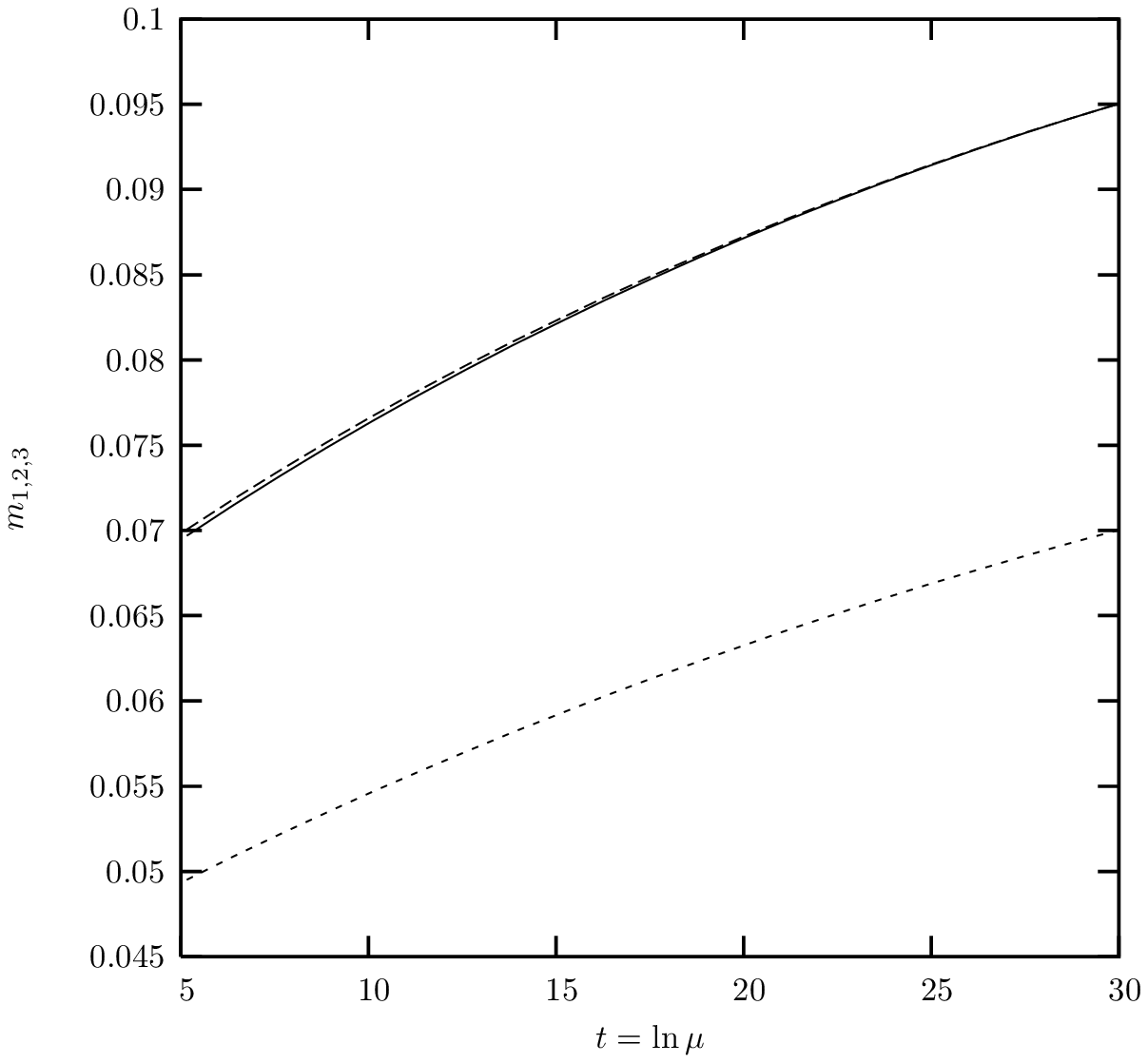}}

{\narrower\narrower\footnotesize\noindent
{Fig.2}
Evolution of the three neutrino mass eigenvalues  in JM conjecture[11]. $m_1$, $m_2$ and 
$m_3$ are represented by  dashed-line, solid-line  and dotted-line respectively.
\par\bigskip}}

\vbox{
\noindent
\hfil
\vbox{
\epsfxsize=10cm
\epsffile [130 380 510 735] {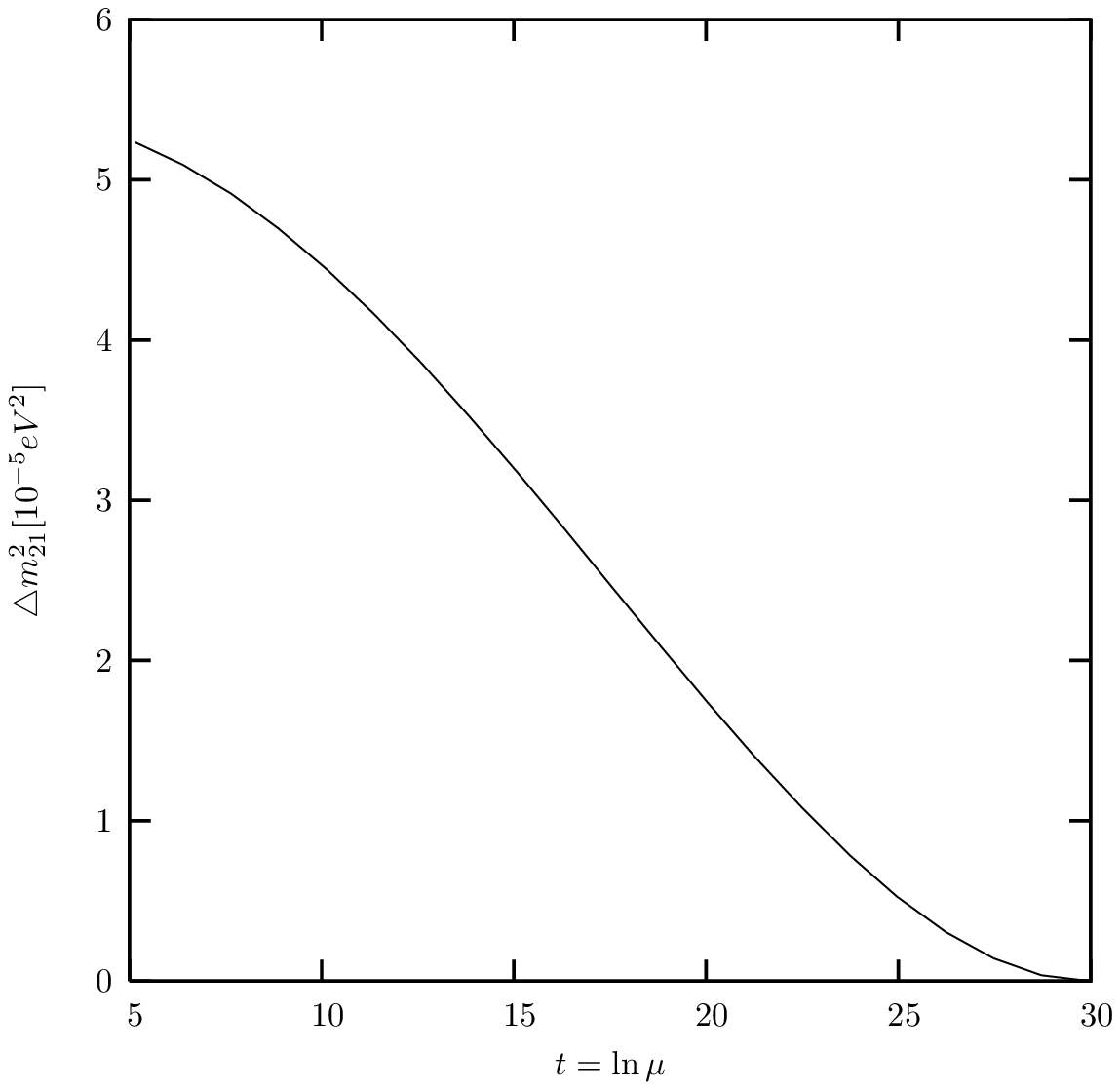}}

{\narrower\narrower\footnotesize\noindent
{Fig.3}
Evolution of the $\bigtriangleup m_{21}^2$ in JM conjecture[11] with energy scale. Its value at high energy scale is zero.
\par\bigskip}}

\vbox{
\noindent
\hfil
\vbox{
\epsfxsize=10cm
\epsffile [130 380 510 735] {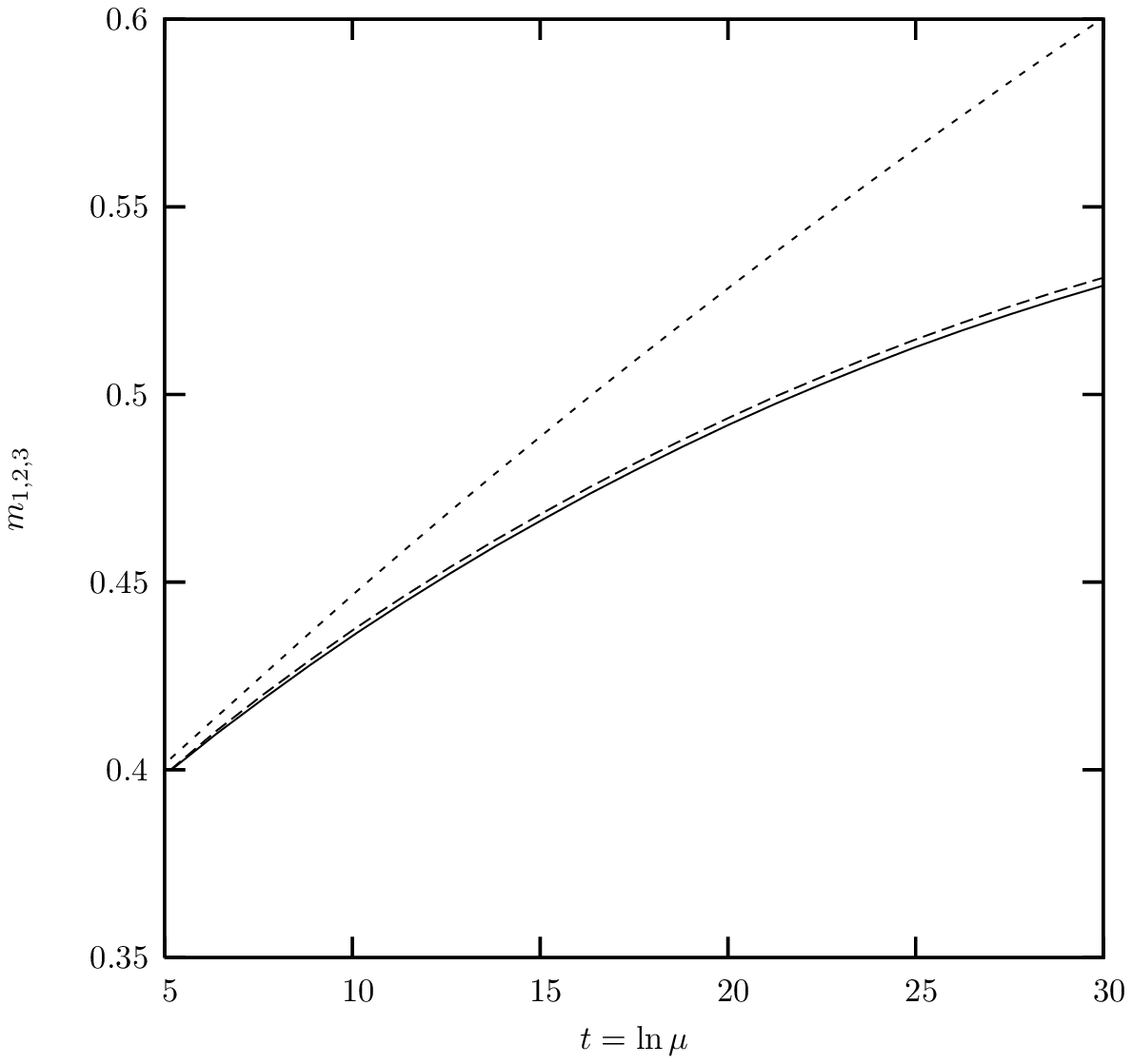}}

{\narrower\narrower\footnotesize\noindent
{Fig.4}
Evolution of the three nearly degenerate neutrino masses  in MPR  conjecture[12] (without the effect of scale-dependent vev in II).
 $m_3$, $m_2$ and $m_1$ 
 are represented by solid line, dashed-line and dotted-line respectively.

\par\bigskip}}

\vbox{
\noindent
\hfil
\vbox{
\epsfxsize=10cm
\epsffile [130 380 510 735] {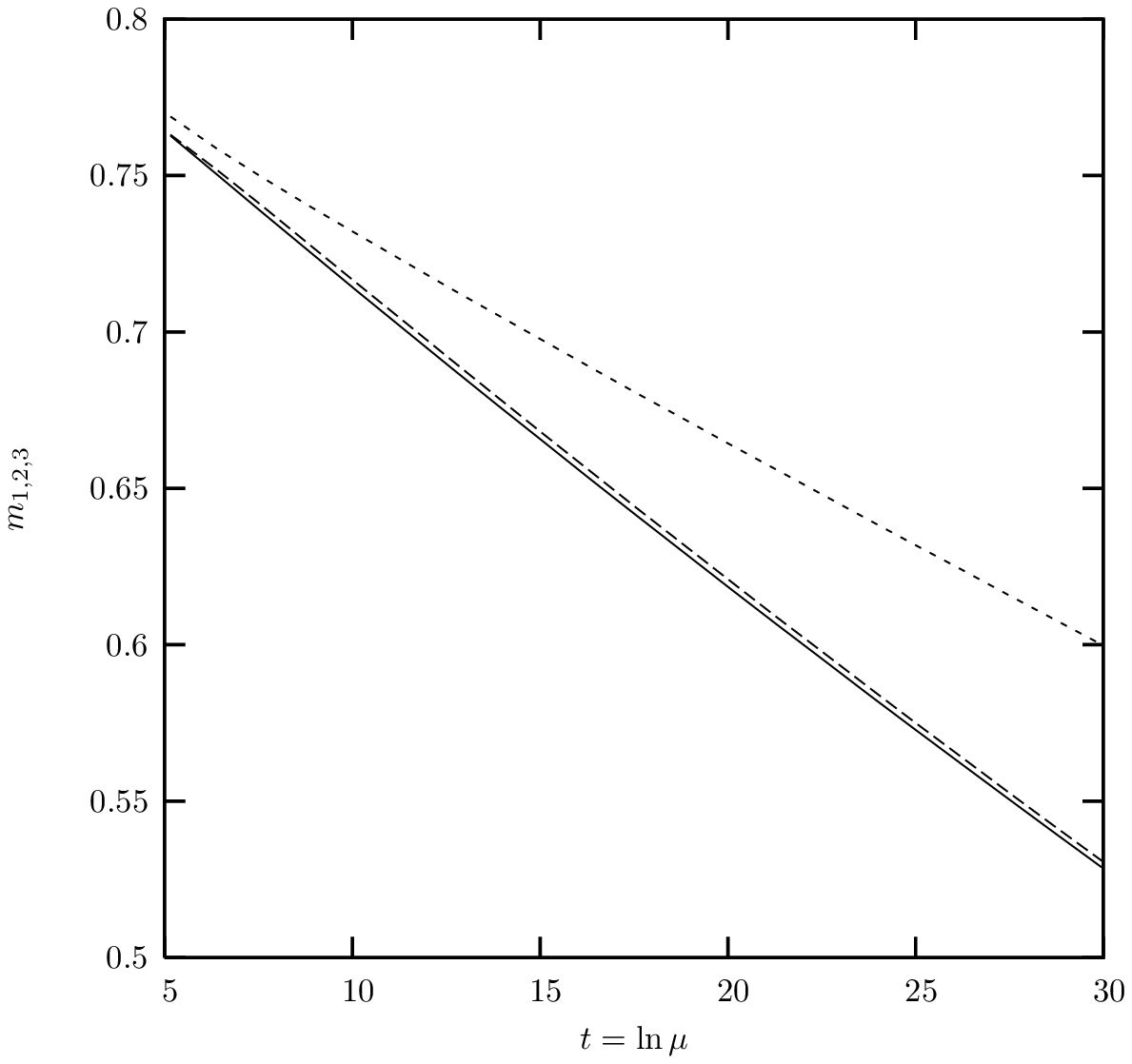}}

{\narrower\narrower\footnotesize\noindent
{Fig.5}
Evolution of the three nearly degenerate neutrino masses  in MPR  conjecture[12] (with the effect of scale-dependent vev in I).
 $m_3$, $m_2$ and $m_1$ 
 are represented by solid line, dashed-line and dotted-line respectively.

\par\bigskip}}

\end{document}